\documentclass[prd,showpacs,preprintnumbers,tightenlines,amssymb,
               eqsecnum,superscriptaddress,floatfix]{revtex4}

\usepackage{graphicx}

\newcommand{\subsubsubsection}[1]
{
  \vspace{12pt}
  {\bf #1}
  \vspace{12pt}
}

\newcommand{\mathinput}[1]
{
  \begin{tt}
  \begin{center}
    #1
  \end{center}
  \end{tt}
}

\newcommand{\mathdialogue}[2]
{
  \begin{center}
  \begin{tabular}[t]{rl}
  {\tt In := } & \parbox{10cm}{\tt {#1}} \\
\\
  {\tt Out = } & \parbox{10cm}{#2} \\
  \end{tabular}
  \end{center}
}

\newcommand{\Tud}[3]
{
  #1 ^#2 _{\phantom{#2} #3}
}

\newcommand{\be}{\begin{equation}}
\newcommand{\ee}{\end{equation}}
\newcommand{\bea}{\begin{eqnarray}}
\newcommand{\eea}{\end{eqnarray}}

\begin{document}
\preprint{gr-qc/yymmnnn}

\title{Kranc: a Mathematica application to generate numerical codes for tensorial
evolution equations}

\author{Sascha Husa}
\affiliation{Max-Planck-Institut fur GravitationsPhysik,
Albert-Einstein-Institut, Am M\"uhlenberg 1, D-14476, Golm, Germany}

\author{Ian Hinder}
\affiliation{School of Mathematics, University of Southampton,
Southampton, SO17 1BJ, United Kingdom}

\author{Christiane Lechner}
\affiliation{Max-Planck-Institut fur GravitationsPhysik,
Albert-Einstein-Institut, Am M\"uhlenberg 1, D-14476, Golm, Germany}

\date{6th April, 2004}

\begin{abstract}
We present a suite of Mathematica-based computer-algebra packages, termed ``Kranc'', 
which comprise a toolbox to convert (tensorial) systems of partial differential evolution equations
to parallelized C or Fortran code. 
Kranc can be used as a ``rapid prototyping'' system for physicists or 
mathematicians handling very complicated systems of partial differential equations,
but through integration into the Cactus computational toolkit
we can also produce efficient parallelized production codes. 
Our work is motivated by the field of numerical relativity,
where Kranc is used as a research tool by the authors.
In this paper we describe the design and implementation of both the Mathematica
packages and the resulting code, we discuss some example applications,
and provide results on the performance
of an example numerical code for the Einstein equations.
\end{abstract}


\maketitle

\section{Introduction}
\label{sec:intro}

The numerical solution of partial differential equations (PDEs) is a
common and often quite challenging
problem in physics, applied mathematics and other fields.
The Kranc software presented here is designed as a tool for PDE systems where
complexity is a principal problem, and manual coding of equations and
general equation-dependent code is too error prone or simply not feasible.
Our software is released under the GPL (GNU General Public License), and is available
at the URL {\tt http://numrel.aei.mpg.de/Research/Kranc}. Currently, Kranc only supports
the numerical solution of initial value problems. Extension to initial boundary value
problems and elliptic boundary value problems is planned for the future.

Our work is motivated by the field of numerical relativity, i.e.~the numerical
solution of the Einstein equations \cite{Luis}.
An important open research problem is to find an evolution
system for the Einstein equations which is well-suited for numerical simulations.
The presented software drastically reduces the complexity of the task of
comparing different formulations of the Einstein equations;
these are complicated nonlinear systems which can have dozens of
evolution variables and thousands of source terms.

We aim to address the problems of handling complexity and of dealing
with tensorial (or other multi-component) quantities in a transparent
way.
To this end, we present a suite of Mathematica packages, termed
``Kranc'', to convert (optionally tensorial) systems of partial
differential evolution equations to parallelized C or Fortran code.
Our approach is to use Mathematica to perform high level tensor
operations and convert the equations into component form.  This form
is then converted to C or Fortran 90.  See Sec.~\ref{sec:role_of_CA} for the rationale
behind this choice.
For evolution problems in three spatial dimensions performance is
critical.  The design decisions that we have made 
(see Sec.~\ref{sec:infrastructure},\ref{sec:codeDesign})
allow us to automatically generate codes with efficiencies comparable
to hand-written code.  Kranc provides a simple programming model for
code generation. Modifying Kranc to generate more efficient code
by--for example--re-arranging loops, is straightforward.


In numerical relativity, the use of computer algebra (CA)
methods based on symbolic tensor calculus can not only save scientists
valuable time, but can also help to substantially reduce errors.
It can be used not only for manipulating the
evolution equations, but also for deriving constraint propagation
systems and perturbation formalisms. Having a code-generation system
directly compatible with the CA system is obviously desirable.
Generally, we feel that our use of CA makes it easier to focus on
algorithms, detached from a particular system of equations. Stressing
a more abstract point of view is not only mathematically more
appealing but also increases flexibility, which benefits scientific
productivity.


Our paper is organized as follows:
In Sec.~\ref{sec:infrastructure} we describe the computational infrastructure that we use.
Since our software only targets the issue of complexity of the equation systems and of 
processing tensorial quantities, we rely on external software to provide
a general computer algebra environment and a framework for parallelization.
The design of the generated codes is described in Sec.~\ref{sec:codeDesign}, while the design
of the Kranc software itself is described in Sec.~\ref{sec:krancDesign}.
In Sec.~\ref{sec:examples} we describe the use of Kranc to generate codes for the Klein-Gordon
equation, the Maxwell equations, and a simple evolution system for general relativity, the ADM
equations. We consider our software both  as a ``rapid prototyping'' system and as
a system to create efficient production code, we therefore also include some results on 
performance.

\section{Choice of Infrastructure, and related design decisions}
\label{sec:infrastructure}

\subsection{The role of computer algebra}
\label{sec:role_of_CA}

When designing a numerical code to solve tensorial equations, two principal 
alternatives come to mind.  A first possibility is to implement high-level tensor operations in
a language like C++ or Fortran 90 using tensorial operations like contractions which operate on
tensor objects. An interesting example of this approach has been 
implemented by Schnetter 
\cite{EriksCode}. The second option is to use computer algebra to expand all tensorial expressions
into components, and then use only standard arithmetics in the numerical code.
Both approaches have their advantages. While the first approach avoids mixing of technologies,
the second approach uses specialized technologies to deal with parts of the problem.

We decided to adopt the second strategy for the following reasons. Most importantly,
we needed a framework that extends beyond code generation, and can also be used  
as a tool for the analytical part of our work. This requirement naturally leads to the use 
of a computer algebra system.
Including the transition from tensors to components and the process of code generation
in the computer algebra system is relatively easy to implement, and greatly simplifies the
structure of the resulting numerical code, which is written purely at the level of components. 
At least part of this task could easily have been implemented in a language like PERL,
which would have decreased the dependence on a commercial computer algebra system (we chose
Mathematica, see Sec.~\ref{sec:infrastructure:CA}). However, this would have demanded 
knowledge of (at least)
three different languages from many users and developers:
 Mathematica, PERL and C/Fortran, which
would have significantly steepened the learning curve.

Our design, where the numerical code is not based on tensorial data structures or
operations, simplifies performance optimization as the code is readable and transparent
down to the level of standard arithmetics. The resulting code can in principle be implemented in 
any suitable language, and we provide the option of generating either Fortran 90 or C
(all examples below will be C).
Furthermore, in our approach the full power of the chosen computer algebra system can be utilized
in the process of code generation, e.g.~to perform optimizations which rely on the structure of the
equations.
Generally, manipulations on the computer algebra level make it much easier to perform experiments involving
significant rewrites of the numerical code.

\subsection{Computer algebra software}
\label{sec:infrastructure:CA}

All of the tasks performed by our software which are related to the tensorial nature of equation
systems can in principle be accomplished by component
calculations.  These can be carried out quite conveniently and efficiently
by computer algebra systems like GRTensorII \cite{grtensor}, which allows
expressions to be entered in an abstract notation and yields
results in component form. This is often what one wants, but when considering 
deriving and analysing different systems of equations, as is often
necessary in numerical relativity research, it is clear that
more is needed.  Apart from the fact that this approach may result in very large calculations
requiring significant time and memory, this method is unwieldy and
not very intuitive.
Rather one would like to keep an abstract notation as long as possible,
and in particular get results in this notation.
We have decided to use abstract index tensor calculus; i.e.~tensors are manipulated 
symbolically according to rules based on an index notation (see e.g.~Wald \cite{Wald},
Sec.~2.4, and Penrose and Rindler \cite{PenroseRindler1984}).

We chose {\em Mathematica} as our base CA system for several reasons.
The authors were already familiar with its use, the language has
proven remarkably stable, and {\em Mathematica} provides excellent
support for pattern matching in the core language.  Also, with the
freely available {\em Ricci} \cite{Ricci}, {\em xTensor} \cite{xTensor} 
and the commercial {\em MathTensor} \cite{MathTensor}, 
there are already three add-on packages
for abstract index tensor manipulation available. 
{\em xTensor} has only been released very recently and was not available
during the main development phase of the present project.
Note that pattern matching seems quite essential for tensor manipulations,
e.g.~$T_{ab}$ and $T_{cd}$ are not the same expression but
still are equivalent mathematical objects, which can easily
be identified with pattern matching techniques. 

Code generation with Kranc can be performed using either {\em MathTensor} or
the TensorTools package, which was developed by I. Hinder during the course of
this work as a non-commercial alternative. 
MathTensor has also been made the basis of our extensive analytical
calculations, but since this article is restricted to the code generation part
of our project, we will discuss only the usage with TensorTools, which is
included in the distribution of Kranc.
Both MathTensor and TensorTools represent tensors as plain functions of
indices:
{\tt h[la, lb]} $\rightarrow$ $h_{ab}$,
{\tt CD[Metricg[la, lb], lc]} $\rightarrow$ $0$. This straightforward syntax
is less error-prone than {\em Ricci}'s corresponding {\tt h[L[a], L[b]]} or {\tt
h[L[a], L[b]] [L[c]]} for a covariant derivative.

TensorTools uses approximately the same syntax as MathTensor, and will
accept reasonably straightforward MathTensor expressions involving
partial derivatives and tensors.  This means that tensorial
analytical work can be done using MathTensor where TensorTools is
inadequate, and the results can be fed into TensorTools and hence
Kranc.

\subsection{Parallel computing infrastructure}
\label{sec:infrastructure:parallel}

Aiming at ``production quality'' evolution codes in three spatial dimensions, 
we regarded parallelization as a vital feature. The primary requirements for the
parallelization strategy were a short development time and
portability -- our code should run both on inexpensive 
small clusters, e.g.~commodity Linux workstations, but should also scale
up to very large runs performed at supercomputer centres. 
We therefore did not want to rely on shared memory or proprietary solutions.
The current de facto standard for distributed (scalable) computing is MPI (message
passing interface) \cite{MPI}.
Several software packages are available which introduce
a software layer between the application programmer and MPI, and thus
significantly reduce the effort to write parallel applications.
Two examples we considered as the basis for our work
are Cactus \cite{Cactus} and PETSc \cite{PETSc}.
PETSc is a general purpose tool developed at the Argonne National Laboratory
as parallel framework for numerical computations, and Cactus 
is an open source problem solving environment originating in the numerical relativity
community, originally developed at the Max Planck Institute for Gravitational Physics.
While PETSc offers more support for numerical algorithms, in particular
for parallel elliptic solvers, Cactus already contains some general numerical
relativity functionality like horizon finders \cite{JonathanAH,PeterEH}
and is geared toward flexibility and
collaborative projects (it also allows the use of PETSc as a library).
We finally decided to base our code on Cactus, also because two of the authors currently work at
the Max Planck Institute for Gravitational Physics,
and thus have a chance of direct collaboration with the Cactus development team.

Cactus mainly targets the issues of parallelization, modularization and portability.
The name Cactus derives from the design of a central core (or "flesh") which connects to 
application modules (or "thorns") through an extensible interface.
Thorns can implement problem specific code on any level from concrete physics applications 
to low-level infrastructure.
A set of thorns comprising the ``Cactus computational toolkit'' provide infrastructure
such as parallel I/O, data distribution and checkpointing. Current data distribution
implementations are based on the principles of domain decomposition and message passing 
using the MPI standard \cite{MPI}.
This choice yields an open and reasonably {\em documented} infrastructure,
parallelization, a variety of I/O methods and allows easy interfacing
with a growing community writing numerical relativity Cactus applications.
Modifications to interface with other systems with capabilities similar to
those of Cactus or with standalone codes should be straightforward.

We now describe those aspects of Cactus which are important for the remainder of this paper.  
Cactus does not have the structure of a library which provides a set of functions that
can be called by the user application. Instead, all Cactus thorns are compiled into libraries,
and management tasks such as code execution and allocation of distributed memory are
handled by Cactus and steered via configuration files. The end user supplies a 
``parameter file'' which specifies all user-controllable aspects of the run.

The basic module structure within Cactus is called a ``thorn''. All user-supplied code
is organized into thorns, which communicate with each other via calls to the flesh API
(application programmer interface) or APIs of other thorns. 
The integration of a thorn into the flesh or with other thorns is specified in
configuration files which are parsed at compile time.
Thorns are logically grouped into {\em arrangements} for organisational purposes.
The arrangements live in the {\tt arrangements} subdirectory of the main
Cactus directory.  
Inside an arrangement directory there are directories for each thorn
belonging to the arrangement. One current shortcoming of Cactus is that this arrangement 
directory structure can only be one level deep. 
Arrangement and thorn names must be (case independently) unique,
must start with a letter, and can only contain
letters, numbers or underscores. In addition, thorn names must contain 27 characters or
less.

A key concept for a thorn is the {\em implementation}, which defines a group of variables
and parameters which are used to implement some well-defined functionality.
Relationships among thorns are all based upon relationships among the
implementations they provide.
Different thorns providing the same implementation can be compiled into the same binary;
the decision for which thorn is executed (``active'') can be made at run time.

A thorn consists of a subdirectory of an arrangement containing at least three
administrative files, illustrated here by examples taken from a Kranc-generated arrangement to
solve the Maxwell equations (see the example in Sec.~\ref{sec:maxwell}).
The complete code is provided with the Kranc distribution.
These files are text files written in the 
{\em Cactus Configuration Language} (CCL).
The CCL syntax is case independent, and the `\#' character indicates that the
rest of the line is a comment.
\begin{itemize}
\item {\tt interface.ccl}\\
This defines the implementation the thorn provides, 
and the variables the thorn needs, along with their
visibility to other implementations.
\begin{verbatim}
# excerpt from Example code EM/EMBase/interface.ccl
implements: EMBase
inherits:   Grid
public:

CCTK_REAL El type=GF timelevels=2 # the electric field
{
  El1, El2, El3                   # 3 components of the vector El
} "El"

CCTK_REAL Elrhs type=GF timelevels=1 # RHS for El-evolution eq. 
{
  El1rhs, El2rhs, El3rhs
} "Elrhs"

# analogous for the magnetic field B ...
\end{verbatim}
\item {\tt param.ccl}\\
This defines the thorn parameters along
with their visibility to other implementations.
The parameters are numbers, strings and switches which are
read at runtime and control the thorn behaviour.
\begin{verbatim}
# excerpt from Example code EM/EMsetID/param.ccl
restricted:
CCTK_REAL sigma "sigma"     # a floating point parameter called sigma
{
  *:* :: "no restrictions"  # all values are allowed 
} 0                         # default value is 0
\end{verbatim}
\item  {\tt schedule.ccl}\\
By default no routine of a thorn will be run. 
Those routines which should be run by the Cactus infrastructure
are registered in the file {\tt schedule.ccl} 
along with directives for appropriate memory allocation and interprocessor
communication for grid variables.
Functions are scheduled in {\em scheduling bins} -- slots in the
main loop such as BASEGRID (for setting up coordinates),
EVOL (the evolution step) or ANALYSIS (for analysing data).
\begin{verbatim}
# excerpt from Example code EM/EMBase/schedule.ccl
schedule EMTTBase_RegisterSymmetries at BASEGRID # BASEGRID is a scheduling bin
{
  LANG: C
} "register symmetries"
\end{verbatim}
\end{itemize}

\section{Design of Kranc Evolution Codes}
\label{sec:codeDesign}

\subsection{Method of Lines}
\label{sec:mol}

Aiming both at simplicity and flexibility, we decided to restrict time evolution
algorithms to those compatible with the method of lines (see e.g.~\cite{mol}).
Here, the discretization of a system of PDEs describing an evolution
problem is split into a separate discretization of ``space''
and ``time''.  Discretizing only the spatial derivatives (e.g.~by finite difference, 
spectral or finite element methods), while keeping the time derivatives
continuous yields a {\em coupled}
set of ordinary differential equations (ODEs) in time, 
with one ODE per spatial grid point. 
These ODEs are coupled via the spatial discretization.
For a stable spatial finite differencing scheme, any stable ODE integrator can
then be used to time-integrate this large system of ODEs.
Further key advantages of the method of lines are that:
\begin{itemize}
\item{the stability theory is well developed, and is simplified by
splitting the analysis into the analysis of spatial and time discretizations;}
\item{numerical methods for the method of lines are well-developed; and}
\item{by decoupling the spatial finite differencing from the
        time integration, the resulting numerical scheme
        and computer code are simplified and can easily be modularized.}
\end{itemize}
In particular, the method of lines makes it straightforward to combine
options for spatial discretization and for time integration.
For the concrete implementation, we choose to base our code on the MoL
\cite{CactusMoL} method of lines thorn within Cactus, developed by Hawke.
This code provides a parallel ODE integrator, implementing 
generic Runge-Kutta and iterative Crank-Nicolson methods.

\subsection{Spatial Discretization}
\label{sec:GenericFD}

Within Kranc, spatial discretization is implemented in a centralized and flexible way.
At the moment, only a small number of finite differencing methods are implemented for
spatial discretization, but extension to other methods based on a regular grid is
straightforward. All finite difference formulas are collected in a single header file,
currently via C-style macros.

For an evolution code in three spatial dimensions, which evolves equations as complicated
as the Einstein equations, performance issues are critical. Is is typical that a code
spends most of its time evaluating the right-hand-side expressions of the
evolution equations. It is thus crucial to simplify these expressions as much as possible
(this is done by the computer algebra part) and optimize the evaluation of the finite
difference formulas. It turns out that significant performance gains can be obtained
when finite difference expressions (and any sub-expressions that are very lengthy or
re-used many times) are promoted to local scalar variables that are pre-computed at the 
start of the loop over grid points. Apart from cutting down runtime, this method
often drastically reduces compile time when heavy optimization is used.
Pre-computing is the default behaviour.

The user of the Kranc system will denote derivatives using an abstract
notation when passing equations to the Kranc functions; e.g.~a first
derivative of the variable $f$ in the x-direction would be denoted as
{\tt D1[f]}.  The complete list of abstract operators for first and
second derivatives is {\tt \{D1, D2, D3, D11, D22, D33, D21, D31, D32\}}.

At compile time, the user has two choices to make:
\begin{itemize}
\item{Should the derivatives be precomputed at the start of the grid
loop?}
\item{What sort of finite differencing should be performed?}
\end{itemize}
In order to disable precomputation, the user should set the
preprocessor macro {\tt NOPRECOMPUTE}, typically through the compiler
option {\tt -DNOPRECOMPUTE}.  The choice of finite differencing is
controlled by defining macros: {\tt FD\_C2} for second order centred, {\tt
FD\_C4} for fourth order centred, or {\tt FD\_C0} for zeroth order
(a projection of all derivative terms to zero,
useful for debugging and performance profiling). These macros may be
defined by compiling with, e.g.~{\tt -DFD\_C2}.  This corresponds
to the default setting.

The GenericFD thorn defines several preprocessor macros for performing
finite differencing.  The concrete finite differencing macro name is
formed by appending the method to the abstract name; for example, {\tt
D1\_c2[f]} denotes the first derivative of $f$ in the x-direction,
using centred second order differences.  At the moment only second
and fourth order centred methods are implemented.

In the C or Fortran code which is produced by Kranc, the abstract
notation is replaced with a grid function style notation; e.g.~{\tt
D1[f]} is replaced by {\tt D1[f,i,j,k]}, where {\tt \{i,j,k\}} are
the loop indices representing the grid points.
These ``derivative operators''
are in fact preprocessor macros, which have different
definitions depending on whether or not precomputation is enabled.
When precomputing is switched off, macros such as {\tt D1[f, i,j,k]}
are defined as {\tt D1gf[f, i,j,k]}.  When precomputing is switched
on, macros such as {\tt D1[f, i,j,k]} are defined as {\tt D1f}. Here
{\tt D1f} is used as a local variable which is pre-computed at the
start of the loop over grid points; e.g.
\begin{center} 
{\tt D1f = D1gf(f,i,j,k);}
\end{center}
The operators with suffix {\tt gf} are defined in terms of whichever
concrete differencing macros have been selected.

\subsection{Boundary Conditions}
\label{sec:boundary_conditions}

Currently, Kranc only fully supports periodic boundary
conditions.  This is achieved via the use of ``ghost points'' on the
sides of the grid which Cactus automatically populates with data from
the opposite side so that finite difference operators can be used
right up to the last ``real'' grid point.

\subsection{Code organisation}
\label{sec:code_organisation}


On the user level, the Kranc code generation toolkit consists of Mathematica functions
that generate an entire code module ({\em thorn} in the Cactus terminology).
Within the Kranc system, thorns are classified into a small number of thorn types,
and complex codes can be composed from a larger number of thorns which fall into
one of these classes.  The basic tasks of a thorn are: 
\begin{itemize}
\item{define Cactus grid functions;}
\item{assign values to grid functions;}
\item{set grid function attributes (e.g.~symmetries or boundary conditions); and}
\item{define Cactus parameters;}
\end{itemize}
We define five types of thorn with specific characteristics regarding their
functionality: the base, setter, translator, evaluator and MoL thorns.
While the base thorn only defines parameters, grid functions and
their properties, all the other thorns also
perform one or more calculations, i.e.~they assign new values to one or more
grid functions.
Here the translator, evaluator and MoL thorns are in a sense ``added value'' versions
of the basic setter thorn.

A {\em calculation} is one of the essential concepts of Kranc. The end
result of a calculation is to assign new values to a grid function in a
loop over grid points. In order to simplify lengthy calculations, all
calculations are carried out in terms of local scalar variables, and
the assignment to grid functions is made at the end of the loop.  This
avoids repeated multiplications due to array accesses.
Intermediate expressions, which are only needed as local scalars 
(i.e.~of which no derivatives are required), are called {\em shorthands}
in the Kranc context.

As an example of a calculation, we list a slightly abbreviated (to save space)
version of the file {\tt MKGMoL\_CalcRHS.c}
(in directory {\tt Examples/KleinGordon/MKG/MKGMoL/src}),
which assigns the right hand sides for the MoL evolution of the 
Klein-Gordon equations.
\begin{verbatim}
// several header files are included and some macros get defined
/* Define macros used in calculations */  // C++ style comments do not appear in original code
#define INITVALUE  (42)

void MKGMoL_CalcRHS(CCTK_ARGUMENTS)
{
  DECLARE_CCTK_ARGUMENTS
  DECLARE_CCTK_PARAMETERS

  /* Declare the variables used for looping over grid points */
  int i = INITVALUE;  // same for j, k, index, istart, jstart, kstart, iend, jend, kend 

  /* Declare finite differencing variables */
  CCTK_REAL dx = INITVALUE;  // same for dy, dz, dxi, dyi, dzi, hdxi, hdyi, hdzi

  /* Declare shorthands */ // none in this example

  /* Declare local copies of grid functions */
  CCTK_REAL phiL = INITVALUE;    // same for phitL, phirhsL, phitrhsL, D11phi, D22phi, D33phi

  /* Initialize finite differencing variables */
  dx = CCTK_DELTA_SPACE(0);
  dxi = 1 / dx;
  hdxi = 0.5 * dxi;  // analogously for x and y directions

  /* Set up variables used in the grid loop with stencils suitable for finite differencing */
  istart = cctk_nghostzones[0];
  iend   = cctk_lsh[0] - cctk_nghostzones[0]; // analogously for jstart, kstart, jend, kend

  /* Loop over the grid points */
  for (k = kstart; k < kend; k++)
  {
    for (j = jstart; j < jend; j++)
    {
      for (i = istart; i < iend; i++)
      {
        index = CCTK_GFINDEX3D(cctkGH,i,j,k); // a Cactus macro to compute the 3D array index

        /* Assign local copies of grid functions */
        phiL = phi[index];
        phitL = phit[index];

        /* Precompute derivatives */
        D11phi = D11gf(phi,i,j,k); // analogous for D22phi and D33phi

        /* Calculate grid functions */
        phirhsL = phitL;

        phitrhsL = D11(phi,i,j,k) + D22(phi,i,j,k) + D33(phi,i,j,k) -
            phiL*SQR(mass);   // use macro SQR for mass * mass

        /* Copy local copies back to grid functions */
        phirhs[index]  = phirhsL;
        phitrhs[index] = phitrhsL;
      }
    }
  }
}
\end{verbatim}

A Cactus arrangement will typically contain one base thorn, which defines
the basic quantities of the model to be studied, and an assortment of
several of the other Kranc thorn types which implement the equations.
Functionality which is required but not presently covered by the code that is generated
by Kranc, e.g.~more involved boundary conditions,
can easily be added as hand coded thorns.
There are currently several restrictions as to what Kranc is able to
generate: only a few boundary conditions are
implemented, and only periodic boundaries are fully tested and
supported.

Note that all thorns can define their own
parameters, which can then be shared with other thorns. Apart from defining
parameters and displaying banner information,
the Kranc thorns implement the following tasks:
\begin{itemize}
\item{Base thorn}:
\begin{itemize}
  \item 
    Register grid functions with Cactus.  Discriminate between
    grid functions that can be evolved and need two sets of storage
    (timelevels) and those that only need one time level allocated.
              
  \item 
    Register grid function symmetries.  The transformation behaviour
    under coordinate reflections $x^i \rightarrow - x^i$ needs to be
    defined in order to use Cactus infrastructure to run in
    bitant/quadrant/octant mode for data which are compatible with
    such symmetries. Via the {\em Cartoon method} \cite{cartoon_paper}
    this infrastructure can also be used to handle
    axisymmetric situations in Cartesian coordinates while keeping
    manifest axisymmetry.
\end{itemize}
\item{Setter thorn}:
\begin{itemize}
        \item Set grid functions to values, e.g.~to set initial data or
              to update auxiliary grid functions.

         Functions are scheduled in the POSTINITIAL and EVOL time bins
         at appropriate times relative to the time stepping functions.
\end{itemize}
\item{MoL evolve thorn}:
\begin{itemize}
  \item
    Register grid functions with the MoL thorn. During
    registration, the grid functions need to be designated as either
    ``evolved'' or ``primitive'', with respect to this thorn.  Evolved
    variables are those that {\em this thorn} will evolve using
    MoL.  Primitive variables are those that are used by this
    thorn in the calculation of the right hand sides of the evolution
    equations, but are not evolved themselves.
  \item
    Generate code to apply boundary conditions.
  \item
    Set MoL right hand side variables through a function scheduled in
    the EVOL bin
\end{itemize}
\item{Translator thorn}:
\begin{itemize}
  \item 
    This is a specialized version of a setter thorn which translates
    between Kranc variables and some other set of variables.  The idea
    is that initial data may be specified in one set of variables by
    an external thorn, and can be translated by a translator thorn
    into the variables you wish to evolve using Kranc.  The evolved
    variables can then be translated back into the original set, so
    that external thorns can use them.  These translations occur in
    the POSTINITIAL and POSTSTEP bins, respectively.

    This interface is essential for compatibility with many of Cactus'
    existing numerical relativity thorns, which use the ADM variables
    via the ADMBase thorn.

\end{itemize}
\item{Evaluator thorn}:
\begin{itemize}
        \item Define new grid functions.
        \item Assign values to new grid functions in routines
              scheduled in the Cactus ANALYSIS bin.
\end{itemize}
\end{itemize}

When defining grid functions, tensor components are grouped together 
as Cactus groups. The reflection symmetries of grid functions
are read off from their tensor properties automatically (through the
function {\tt calcSymmetry[gf]} in file {\tt Thorn.m}). By default,
grid functions are treated as scalars,  but e.g.~a grid function
called h11 is interpreted as the 11-component of a 2-tensor, which
defines its parity information.

\section{Kranc Design}
\label{sec:krancDesign}

Kranc is composed of several Mathematica packages (textual .m files);
each one performs a distinct function.
The user only needs to be concerned
with calling functions from one package: KrancThorns.  This package
contains functions for creating the different types of Kranc thorn.

\begin{figure}
\label{fig:kranc_design}
\includegraphics[clip,width=16cm]{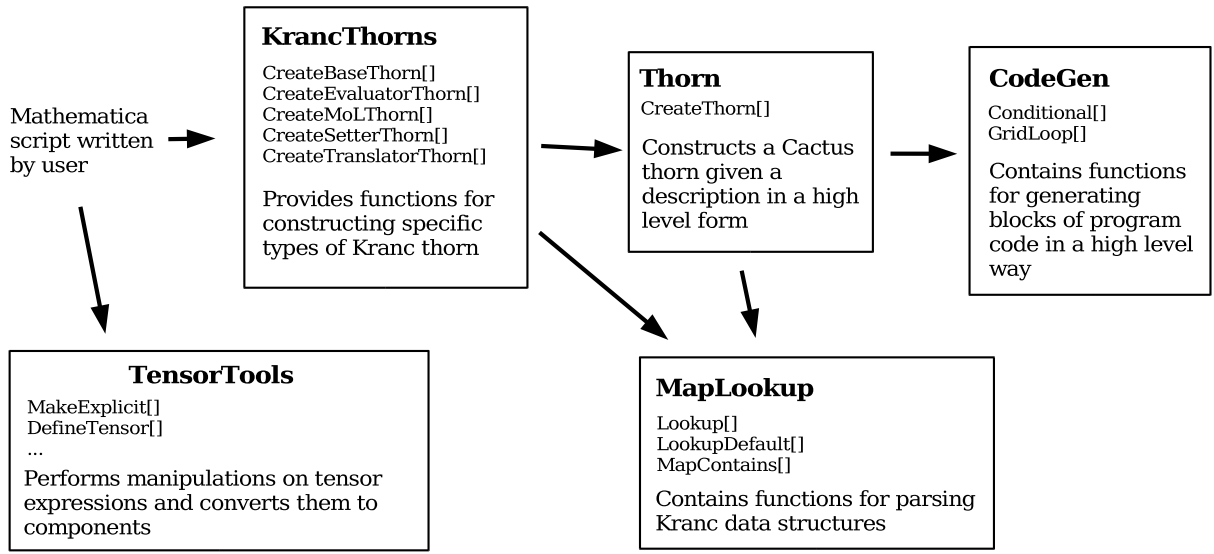}
\caption{Relationships between Kranc packages: 
Each block represents a package, with the main functions it provides
indicated with square brackets.  An arrow indicates that one package
calls functions from another}
\end{figure}

The diagram in Fig. \ref{fig:kranc_design}
illustrates the relationships between the Kranc packages
KrancThorns, TensorTools, CodeGen, Thorn and MapLookup, which
are described in the following subsections. 
Structuring the code in this way, separating out the different
logically independent functions, promotes code reuse.  For example,
none of the thorn generation packages need to know anything about
tensors, and none of the packages other than CodeGen (in principle)
need to know anything about the programming language that is being
written (C or Fortran).  We have chosen to define the concept of a
``setter thorn'', and ``evaluator thorn'' etc, but the mechanics of
producing a thorn implemented in Thorn and CodeGen are completely
independent of this decision.


\subsection{KrancThorns: Constructing the different Kranc thorns}

We have already discussed the different types of Cactus thorn used in
a Kranc arrangement (MoL, setter, base, translator and evaluator). The
KrancThorns package provides functions to create thorns of these types
given high level descriptions.
These  {\tt Create*Thorn} functions are the ones directly called by users;
see the appendix (\ref{app:function_reference}) 
for detailed argument descriptions.
Internally the KrancThorns package uses the Thorn package to create the Cactus
thorns.

\subsubsection{Types of arguments}

Mathematica allows two types of arguments to be passed to a function.
``positional arguments'' and ``named arguments'' (referred to in the
Mathematica book as ``optional arguments''.  It is possible for some
named arguments to be omitted from a function call; in this case a
suitable default will be chosen.  Positional arguments are useful when
there are few arguments to a function, and their meaning is clear in
the calling context.  Named arguments are preferred when there are
many arguments, as the argument names are given explicitly in the
calling context.

For each type of Kranc thorn, there is a function to create it ({\tt
Create*Thorn}).  There is a certain set of named arguments (``Common
named arguments'') which can be passed to any of these functions
(e.g.~the name of the Thorn to create, where to create it, etc).
Then, for each type of thorn, there is a specific set of named
arguments specifically for that thorn type.  All of the functions
accept some positional arguments as well.
See the appendix (\ref{app:function_reference}) for descriptions of the
arguments that can be given to each of the KrancThorns functions.

\subsubsection{Common data structures}

Kranc is a relatively complex system.  It consists of several packages
which need to pass data between themselves in a structured way.
Mathematica does not have the concept of a C++ ``class'' or a C
``structure'', in which collections of named objects are grouped
together for ease of manipulation.  Instead, we have adopted the
definition of
a ``Kranc structure'' as a list of rules of the form {\tt {\it key} ->
{\it value}}.  We have chosen to use the ``rule'' symbol ``{\tt ->}''
for syntactic convenience.
For example, one might describe a person using a ``Person'' structure
as follows:
\begin{center}
\begin{tt}
alice = \{Name -> "Alice",
         Age -> 20,
         Gender -> Female\}
\end{tt}
\end{center}
Once a structure has been built up, it can be parsed with the {\tt
lookup} function in the MapLookup package.  {\tt lookup[structure,
key]} returns the value in {\tt structure} corresponding to {\tt key}.
For example, {\tt lookup[alice, Age]} would return the number 20.
This usage mirrors what is known as an association list (or alist) in
LISP style languages.
Based on this concept we have defined a number of data structures which will
be used to
describe the thorns to construct.  Each of these data structures is
introduced below, and full details are given in the appendix.

\subsubsubsection{GroupDefinition}

A {\tt GroupDefinition} structure lists the grid functions that are
members of a specific Cactus group.  A list of such structures should
be supplied to all the KrancThorns functions so that Kranc can
determine which group each grid function belongs to.

\subsubsubsection{Calculation}

Calculation structures are the core of the Kranc system; a detailed specification
is given in appendix (\ref{app:Calculation}). The idea is that
the user provides a list of equations of the form ${\it variable} \to
{\it expression}$.  When the calculation is performed, for each point
in the grid, {\it expression} is evaluated and placed into the grid
function {\it variable}.
Here {\it expression} may contain partial derivatives of grid functions
written using the following notation:
\begin{displaymath}
\frac {\partial g} {\partial x^i} \to \mbox{\tt Di[g]}
\qquad
\frac {\partial g} {\partial x^i \partial x^j} \to \mbox{\tt Dij[g]}
\end{displaymath}
where $i$ and $j$ must be given explicitly as integers 1, 2 or 3
representing $x$, $y$ or $z$.

You may specify intermediate (non-grid) variables called
``shorthands'' which can be used as {\it variable} for precomputing
quantities which you will use later in the calculation.
To identify these variables as shorthands, they should be listed in a
``Shorthands'' entry of the Calculation.

The arrangement of the terms in the equations can have a marked effect
on both compile time and run time.  It is often helpful to tell
Mathematica to collect the coefficients of certain types of term,
rather than expanding out entire expressions.  To this end, the user
can include a ``CollectList'' entry in a calculation; this is a list
of variables whose coefficients should be collected.

There is the facility for performing multiple loops in a single
calculation structure; this can be used to set a grid function in one
loop, then evaluate derivatives of it in a later loop.  For this
reason, the equations are given as a list of lists of equations.
Explicit synchronization of ghost zones is performed after each loop
for those groups whose grid functions have been set.

Note that the system is not designed to allow the same grid function
to be set more than once in a single loop of a calculation.

\subsubsubsection{GroupCalculation}

A {\tt GroupCalculation} structure associates a group name with a {\tt
Calculation} which is used to update the grid functions in that
group. This is used when creating evaluator thorns, where the
calculations are triggered by requests for output for specific groups.


\subsection{TensorTools}

\subsubsection{Overview}

The TensorTools package was written specifically for the Kranc system,
though it is in no way tied to it.  We needed to perform certain
operations on tensorial quantities, and there was no free software
available which met our needs.
TensorTools has the following features:
\begin{itemize}
\item{Expand covariant derivatives in terms of partial derivatives and
Christoffel symbols (multiple covariant derivatives can be defined)}
\item{Expand Lie derivatives in terms of partial derivatives}
\item{Automatic relabelling of dummy indices to avoid conflicts}
\item{Convert abstract tensor expressions into components}
\end{itemize}

Before using any TensorTools functions, the TensorTools package must
be loaded.  It is self-contained, with no dependencies on any other
packages.

\mathinput{Get["TensorTools.m"];}

\subsubsection{Representation of tensor quantities}

Tensorial expressions are entered in mostly the same syntax as is used
by MathTensor.  An abstract tensor consists of a {\em kernel} and an
arbitrary number of abstract {\em indices}, each of which can be {\em
upper} or {\em lower}.  Abstract indices are alphabetical characters
(a-z, A-Z) prefixed with either an l or a u depending on whether the
index is considered to be lower or upper.  The tensor is written using
square brackets as
\begin{center}
\begin{tt}
kernel [ indices separated by commas ]
\end{tt}
\end{center}
For example, $T_a^{\phantom{a}b}$ would be written as {\tt T[la,ub]}.
There is no automatic index raising or lowering with any metric.
Before using a tensor, it must be registered with the TensorTools
package using the {\tt DefineTensor} function:
\mathinput{DefineTensor[T]}
Entering a tensorial expression causes it to be rendered in an easy to
read form:
\mathdialogue{T[la,lb]}{$T_{ab}$}
Internally, tensors are represented as {\tt Tensor[{\it kernel},
TensorIndex[{\it label}, {\it type}], ...]} where {\it label} is the
alphabetical index, and {\it type} is either ``u'' or ``l'' depending
on the position of the index.  This representation helps in pattern
matching, and allows TensorTools to identify whether a certain object
is a tensor or not.

\subsubsection{Expansion of tensorial expressions into components}

The function {\tt MakeExplicit} converts an expression containing abstract
tensors into a list of component expressions:
\begin{center}
  \begin{tabular}[t]{rl}
  {\tt In := } & \parbox{10cm}{\tt MakeExplicit[T[la, lb]g[ub, uc]]} \\
\\
  {\tt Out = } &  \begin{tabular}[t]{rll}
\{ & g11 T11 & + g21 T12 + g31 T13, g12 T11 + g22 T12 + g32 T13, \\
   & g13 T11 & + g23 T12 + g33 T13, g11 T21 + g21 T22 + g31 T23, \\
   & g12 T21 & + g22 T22 + g32 T23, g13 T21 + g23 T22 + g33 T23, \\
   & g11 T31 & + g21 T32 + g31 T33, g12 T31 + g22 T32 + g32 T33, \\
   & g13 T31 & + g23 T32 + g33 T33\}\\
  \end{tabular} \\
  \end{tabular}
  \end{center}
Note here that upper and lower indices are not distinguished between
in the component form.  TensorTools was written mainly for automated
code generation rather than symbolic manipulation; we suggest using
different kernels for the different forms if this is a problem.

\subsubsection{Covariant derivatives}

TensorTools allows the user to define as many covariant derivatives
(and associated Christoffel symbols) as are necessary.  The following
defines a covariant derivative operator {\tt CD} with Christoffel
symbol {\tt H}:
\mathinput{DefineConnection[CD,H]}
The function {\tt CDtoPD} is used to replace covariant derivatives
with partial derivatives in any expression:
\mathdialogue
{
  CDtoPD[CD[V[ua],lb]]
}
{
  $V^a,\,_b + H^a_{\phantom{a}bc} V^c$
}
The function {\tt MakeExplicit} will automatically do this before
converting expressions into components.

The fundamental operation is to convert a first covariant derivative
of an abstract tensor into a partial derivative plus some extra terms
involving Christoffel symbols.  For convenience, TensorTools
understands that covariant derivatives are linear and satisfy the
Leibniz property.  Also, higher order covariant derivatives are
converted to repeated application of a first order derivative.

We define a number of rules to perform the operations in the reduction
to the desired form.  In the following, $x$ and $y$ represent
expressions which may or may not contain tensorial indices.

\begin{itemize}
\item{Replace any high order covariant derivatives with repeated
application of a first order covariant derivative.  This ensures that
we only need to know how to evaluate a first derivative.
$$\nabla_d \nabla_a V^b \to \nabla_d ( \nabla_a V^b)$$}
\item{Replace the covariant derivative of a product using the Leibniz
rule: $$ \nabla_a (x y) \to (\nabla_a x) y + x (\nabla_a y) $$}
\item{Replace the covariant derivative of a sum using the linearity
property: $$ \nabla_a (x + y) \to \nabla_a x + \nabla_a y $$}
\item{Replace the covariant derivative of an arbitrary expression
containing tensorial indices with its expansion in terms of a
partial derivative and Christoffel symbols, one for each
index in the expression:  e.g.~$$\nabla_a V^b \to \partial_a V^b +
\Gamma^{b}_{\phantom{b}ac} V^c$$ }
\end{itemize}

\subsubsection{Lie derivatives}

The Lie derivative of an expression $x$ with respect to a vector $V$ is
written
\mathinput{Lie[x,V]}
where $V$ has been registered using {\tt DefineTensor} and is written {\em without}
indices.  The function {\tt LieToPD} is used to replace Lie derivatives
with partial derivatives:
\mathdialogue
{
  LieToPD[Lie[T[ua,lb], V]]
}
{
  $\Tud T a {b,c} V^c + \Tud T a c \Tud V c {,b} - \Tud T c b V^a_{,c}$
}
Lie derivatives of products and sums are supported.
The function {\tt MakeExplicit} will automatically perform this replacement
before converting expressions into components.

\subsubsection{Automatic dummy index manipulation}

When two expressions both containing a dummy index $b$ are multiplied
together, one dummy index is relabelled so as not to conflict with any
other index in the resulting expression:

\mathdialogue
{(T[la, lb]g[ub, uc])v[ub, ld, lb]}
{$T_{ab} g^{bc} V^e_{\phantom{e}de}$}
This requires that every multiplication be checked for tensorial
operands.  This can be a performance problem, so the feature can be
enabled and disabled with {\tt SetEnhancedTimes[True]} and {\tt
SetEnhancedTimes[False]}.  It is enabled by default.


\subsection{CodeGen: High level code generation}

During the development of the Kranc system, we explored two different
approaches to generating Cactus files using Mathematica as a
programming language.  Initially, we used a very straightforward
system whereby C statements were included almost verbatim in the
Mathematica script and output directly to the thorn source file.  This
approach has two main deficiencies:
\begin{itemize}
\item{The same block of text might be used in several places in the
code.  When a bug is fixed in one place, it must be fixed in all.}
\item{It is not easy to alter the language that is produced--for
example, it is difficult to output both C and Fortran.}
\item{The syntax in the Mathematica source file is ugly, with lots of
string concatenation, making it difficult to read and edit}.
\end{itemize}

To address the first problem, we decided to create Mathematica
functions to represent each block of text.  This also allows the block
to be customized by giving the function arguments.  By making this
abstraction, it became very easy to change between outputting C and
Fortran, and the resulting Mathematica scripts are more elegant. We
call this system ``CodeGen''.
The effect of migrating our old code \cite{ERE}
to the CodeGen system was that the Mathematica source
files became dramatically shorter and easier to maintain.

Fundamental to the CodeGen system is the notion of a CodeGen ``block'';
this can be either a
string or a list of CodeGen blocks (this definition is recursive).
All the CodeGen functions return CodeGen blocks, and the lists are all
flattened and the strings concatenated when the final source file is
generated.  This is because it is syntactically easier in the
Mathematica source file to write a sequence of statements as a list
than to concatenate strings.

Many programming constructs are naturally block-structured; for
example, C {\tt for} loops need braces after the block of code to loop
over.  For this reason, it was decided that CodeGen functions could
take as arguments any blocks of code which needed to be inserted on
the inside of such a structure.

By default, Kranc generates C code.  To generate Fortran code
the command {\tt SetSourceLanguage{"Fortran"}} has to be issued
after loading {\tt KrancThorns.m}.
Note that while Kranc-generated C-code should always compile (if
different thorns have been created consistently), a Kranc generated
Fortran code may not compile in certain cases. The reasons are
that the Fortran standard limits names to 31 characters, and specifies
a maximum  number of 19 continuation lines per statement
(typical compilers have much larger continuation line limits).
Both limits can be compiler dependent, and are not checked by Kranc.

\subsection{Thorn: Constructing the most general Cactus thorn}

We have decided to define several different types of Kranc thorn
(base, MoL, setter, evaluator, translator) and to provide convenient
interfaces for generating each of them.  However, there is a lot that
is common to each of these, and it would be wasteful for the
underlying implementation to have separate code for generating each
type of thorn.
The Kranc system is designed to be as modular as possible, with as
much code re-use as possible.  To this end, there is a package called
``Thorn'' which takes as input a high level description of a Cactus
thorn, and which creates the necessary thorn files.

The interface.ccl, schedule.ccl and param.ccl files are first created
by calling {\tt CreateInterface}, {\tt CreateParam} and {\tt
CreateSchedule} with high-level descriptions of these files, which are
converted into CodeGen blocks representing the CCL files.  These
blocks are then passed, along with blocks for each source file, to
{\tt CreateThorn}, which creates the final thorn.
The Thorn package provides every aspect of thorn generation that is
not specific to the Kranc thorns we have defined, and could be used
for creating any type of thorn.


\section{Examples}
\label{sec:examples}
In order to illustrate how the packages TensorTools.m and KrancThorns.m 
can be used, we give three examples. 
The first example is the massive Klein Gordon equation
which, although very simple, contains the essential steps for automatic thorn
generation. It is described in some detail 
to serve as a reference for analogous steps in the other 
examples. As a second (still trivial) example, we consider the 
vacuum Maxwell equations. Here the evolution variables are 
components of tensors and therefore the concept of groups and a 
{\tt GroupDefinition} structure becomes essential.
Finally, as a non-trivial example, we consider the Einstein
equations in the standard ADM form.
The automatic generation of Cactus thorns for these equations
involves (almost) the full functionality of the packages TensorTools.m and 
KrancThorns.m.

The following descriptions should be understood as accompanying 
explanations of the scripts {\tt Examples/KleinGordon/MKGTT.m}, 
{\tt Examples/Maxwell/EMTT.m} and {\tt Examples/ADM/KrancADMTT.m}   
which actually perform the automatic generation of code.
The reader therefore is strongly advised to read the text along with the
example scripts.  Detailed instructions for using the examples are
provided in the file README.

\subsection{The Massive Klein Gordon Field}
As a simple example we consider the massive Klein Gordon field equation
\be\label{eq::MKG}
\square \phi - m^2 \phi = 0,  
\ee
where $\square$ is the wave operator on flat space, 
$\square \equiv -\partial^2_t + \delta^{ij} \partial_{i} \partial_{j}$ in
standard Cartesian coordinates $(t, x^i)$.

\subsubsection{Continuum Formulation}

Introducing $\Pi \equiv \partial_t \phi$ reduces Eq.~(\ref{eq::MKG})
to a system of PDEs which are first order in time
\bea\label{eq::MKG1storder}
\partial_t \phi &= &\Pi, \nonumber\\
\partial_t \Pi & = & \delta^{ij} \partial_i \partial_j \phi - m^2
\phi.
\eea

We consider Eqs.~({\ref{eq::MKG1storder}}) on a 3-torus $\mathcal T$ 
(i.e.~on a
rectangular domain with periodic boundary conditions), with initial data
\be\label{eq::MKGID}
\phi (0, x^i) = \cos( 2\pi x/d), \qquad \Pi (0,x^i) = 0,
\ee
where $d$ is the range of the coordinate $x$, $-d/2 \le x \le d/2$.

As an example of a quantity which is evaluated for analysis purposes,
we introduce the total energy $E = \int_{\mathcal T} \rho \, \mbox{d}^3x$, 
with energy density $\rho$,
\be\label{eq::MKGendens}
\rho = \Pi^2 + \delta^{ij} \partial_i \phi \partial_j \phi + m^2 \phi^2.
\ee
The energy is conserved in time and can be used to monitor the accuracy
of the numerical solution.

\subsubsection{Specifications for Kranc}

The script {\tt MKGTT.m} \, in the directory {\tt Examples/KleinGordon}
generates Cactus thorns to solve this time evolution problem.
After loading the packages TensorTools.m and KrancThorns.m
we specify Eqs.~(\ref{eq::MKG1storder}) and 
(\ref{eq::MKGendens}) as rules written in TensorTools syntax; i.e.
\begin{verbatim}
EvolEqs = {dot[phi]  ->  phit,
           dot[phit] ->  PD[phi,la,lb] KD[ua,ub] - mass^2 phi},
\end{verbatim}
where {\tt KD} denotes TensorTools' Kronecker delta and the variable $\Pi$
has been renamed to {\tt phit} in order to avoid conflicts with the 
constant $\pi$ in a Fortran code.
The sums over dummy indices are expanded automatically 
when applying the function {\tt MakeExplicit}, 
\begin{verbatim}
{dot[phi]  -> phit,
 dot[phit] -> D11[phi] + D22[phi] + D33[phi] - mass^2*phi}
\end{verbatim}

The numerical implementation of the above problem is as follows.
The evolution variables {\tt phi, phit},
and the parameter {\tt mass} are declared in a base thorn.
Initially the grid functions {\tt phi, phit} are set to the values 
(\ref{eq::MKGID}) by a setter thorn. The right hand sides of
the evolution equations
(\ref{eq::MKG1storder}) are specified in the MoL thorn.
Finally the energy density {\tt rho} (\ref{eq::MKGendens}) is evaluated
by an evaluator thorn. 

The evolution variables {\tt phi} and {\tt phit} have to be
given in form of groups 
\begin{verbatim}
MKGvars = {{"phi", {phi}}, {"phit", {phit}}};
\end{verbatim}
as arguments to the functions 
{\tt CreateBaseThorn}, {\tt CreateMoLThorn}, {\tt CreateSetterThorn}
and should also be given to {\tt CreateEvaluatorThorn}.
There is no need to introduce any primitives and therefore we set
\begin{verbatim}
MGKprimitives = {};
\end{verbatim}
which is given as a positional argument to the function
{\tt CreateBaseThorn}.

The parameter {\tt mass}  
\begin{verbatim}
MKGRealBaseParameters = {mass};
\end{verbatim}
is declared in the base thorn and is used in the MoL thorn and the
evaluator thorn. It is listed in the  {\tt RealBaseParameters}
argument to the functions {\tt CreateBaseThorn}, {\tt CreateMoLThorn} and
{\tt CreateEvaluatorThorn}. 

Equations (\ref{eq::MKG1storder}), (\ref{eq::MKGID}) and
(\ref{eq::MKGendens}) and the corresponding grid functions 
are specified in calculations. For the evolution equations we define 
\begin{verbatim}
MoLCalculation = {Equations -> {{dot[phi]  -> phit, 
                                 dot[phit] -> D11[phi] + ...}}};
\end{verbatim}
Note that the equations are given as a list of lists (consisting
of one element here), each sublist resulting in a separate 
loop in the code. 
 
The calculations {\tt endensCalculation} for the energy density and
{\tt IDCalculation} for the initial data are set up in an analogous way. 
In addition we assign the name {\tt Name -> "time\_symmetricID"} 
to the function that sets the initial data. 
The function {\tt "time\_symmetricID"} in the Setter thorn should be
called only at the beginning, 
which is achieved by {\tt SetTime -> "initial\_only"}. 

The function {\tt CreateEvaluatorThorn} takes two positional arguments,
{\tt EvaluationDefinitions} and a group structure 
which should contain the groups to be evaluated by the thorn (as well as all
the groups who's variables are referred to). 
{\tt EvaluationDefinitions} specifies the calculation {\tt endensCalculation}
which is responsible for updating the variables in the group {\tt "endens"},
\begin{verbatim}
EvaluationGroups = {{"endens", {rho}}};
EvaluationDefinitions = {{"endens", endensCalculation}};
\end{verbatim}

We also specify a name for the arrangement,
\begin{verbatim}
MKGSystemName = "MKG";
\end{verbatim}
Summarizing, the following information is given to the functions
{\tt Create*Thorn}:
\begin{verbatim}
baseThornInfo = CreateBaseThorn[Join[MKGvars, MKGprimitives],
    Map[First, MKGvars],
    Map[First, MKGprimitives],
    SystemName         -> MKGSystemName,
    RealBaseParameters -> MKGRealParameters];

mainMoLThornInfo = CreateMoLThorn[MoLCalculation, MKGvars,
    RealBaseParameters -> MKGRealParameters,
    PrimitiveGroups    -> Map[First, MKGprimitives], 
    SystemName         -> MKGSystemName,
    ThornName          -> MKGSystemName <> "MoL",
    DeBug              -> True];

setIDThornInfo = CreateSetterThorn[IDcalculation,MKGvars,
    SystemName -> MKGSystemName,
    ThornName  -> MKGSystemName <> "setID",
    SetTime    -> "initial_only",
    DeBug      -> True];  

evaluateThornInfo = CreateEvaluatorThorn[EvaluationDefinitions,
    Join[EvaluationGroups, MKGvars],
    RealBaseParameters -> MKGRealParameters,
    SystemName         -> MKGSystemName,
    ThornName          -> MKGSystemName <> "evalEnergydens",
    DeBug              -> True];
\end{verbatim}
 
As a last step, a thorn list {\tt MKG.th} 
is created by the function {\tt CreateThornList},
\begin{verbatim}
thorns = Join[baseThornInfo, mainMoLThornInfo, setIDThornInfo, evaluateThornInfo];
CreateThornList[thorns, SystemName -> MKGSystemName];
\end{verbatim}

The parameter file for running the code has to
activate all necessary thorns.  In this case, all the thorns in the file 
{\tt MKG.th} are necessary.  The syntax is
\begin{verbatim}
ActiveThorns = "Boundary CoordBase ..."
\end{verbatim}
The parameter {\tt mass} is specified as
\begin{verbatim}
MKGBase::mass = 1.0
\end{verbatim}

We refer the reader to the Cactus thorn documentation for details of
the many standard parameters which can be adjusted, for example to set
up the grid in a different way, or to change the MoL time integration
scheme.  Here we just mention the parameters
necessary for changing the resolution and the Courant factor ($\Delta t / \Delta x$).
The number of grid points in the $x$-direction is set by
{\tt driver::global\_nx}. This number includes both the physical grid
points plus the ghost points,
which are specified by {\tt pugh::ghost\_size}. 
{\tt grid::dxyz} gives the grid spacing (the same for $x$, $y$ and $z$). {\tt cactus::cctk\_initial\_time}
and {\tt cactus::cctk\_itlast} set the initial time and the last time step.
The Courant factor is specified by {\tt time::dtfac}.
 
\subsection{The Maxwell Equations}
\label{sec:maxwell}

As a second example we consider the vacuum Maxwell equations
on a Minkowski background in standard Cartesian coordinates $(t, x^i)$, 
\be\label{eq::EM4d}
\partial_{\mu} F^{\mu \nu} = 0, \qquad  \partial_{[\mu} F_{\nu \sigma]} = 0,
\ee 
where $F_{\mu \nu}$ is the electromagnetic field strength.

\subsubsection{Continuum Formulation}

An observer at rest with respect to the  
above coordinates measures an electric and magnetic
field
\be
E_i = F_{i 0}, \qquad B_i = \epsilon_i{}^{kl} F_{k l}, 
\ee
where $i, k, l$ run from  $1$ to $3$ and $\epsilon_{ijk}$ is the 
alternating symbol with $\epsilon_{123} = 1$.
Eqs.~(\ref{eq::EM4d}) can be decomposed into the evolution equations
\bea\label{eq::EMEvolEqs}
\partial_t E_i & =& \epsilon_{i}{}^{jk} \partial_j B_{k}\nonumber\\
\partial_t B_i & =& - \epsilon_{i}{}^{jk} \partial_j E_{k}
\eea
and the constraints
\be\label{eq::EMConstr}
C_E \equiv \delta^{ij} \partial_{i} E_{j} = 0, \qquad C_B
\equiv \delta^{ij} \partial_i B_{j}  = 0.
\ee
If the constraints (\ref{eq::EMConstr}) are satisfied initially 
the evolution equations (\ref{eq::EMEvolEqs}) guarantee that
they are satisfied for later times as well. It is therefore sufficient
to solve (\ref{eq::EMEvolEqs}) subject to initial data
satisfying (\ref{eq::EMConstr}).

Again we take the spatial domain to be a 3-torus $\mathcal T$ 
and consider
the initial data
\bea\label{eq::EMID} 
& &E_1 (0, x^i) = \sigma \cos( \frac{2\pi (x+y)}{d}), \nonumber\\
& &E_2(0,x^i) = (\sigma -1) \cos(\frac{2 \pi x}{d}) - 
                \sigma \cos(\frac{2\pi (x+y)}{d}), \nonumber\\
& &B_3(0,x^i) = (1-\sigma)\cos (\frac{2 \pi x}{d}) + \sigma \cos(\frac{2\pi
(x+y)}{d}) \nonumber\\
& & E_3(0,x^i) = B_1(0, x^i) = B_2(0,x^i) = 0,
\eea
where $-d/2 \le x,y \le d/2$.
For $\sigma = 0$ these data evolve as a wave travelling in $x$ direction,
$E_2(t,x^i) = - \cos(\frac{2\pi (x +t)}{d}) = -B_3(t, x^i)$, with all other
components vanishing. For $\sigma = 1$ the wave evolves in
a diagonal direction in the $xy$ plane.

The energy density $\rho$ is given by
\be
\rho = \frac 1 2 \delta^{ij} ( E_i E_j + B_i B_j ).
\ee
and gives rise to the total energy 
$E = \int_{\mathcal T} \rho \, \mbox{d}^3x$, which again is conserved in time.

\subsubsection{Specifications for Kranc}

The procedure for generating Cactus thorns is similar to the previous
example, 
the major difference being that the variables are components
of tensors and equations (\ref{eq::EMEvolEqs}) have to be split 
into equations for the individual components.

The automatic generation of code is performed by
{\tt Examples/Maxwell/EMTT.m}.
We define the tensors {\tt El} and {\tt B} with TensorTools' function  
{\tt DefineTensor}, 
where the electric field has been renamed in order to avoid a name clash
with the exponential constant in Mathematica.
The evolution equations (\ref{eq::EMEvolEqs}) in TensorTools syntax read
\begin{verbatim} 
EvolEqs = {{dot[El[la]] ->  (Eps[la,lb,lc] KD[ub,ue] KD[uc,uf] PD[B[lf],le]),
            dot[B[la]]  -> -(Eps[la,lb,lc] KD[ub,ue] KD[uc,uf] PD[El[lf],le])}};
\end{verbatim}
These equations are split 
into individual components by applying the function {\tt MakeExplicit}.

As before, we generate a 
base thorn, a MoL thorn, a setter thorn for the initial data and an 
evaluator thorn for evaluating the constraints and energy density.

The evolution variables are the electric and magnetic field
\begin{verbatim}
EMvars = {{"El", {El1, El2, El3}}, {"B", {B1, B2, B3}}};
\end{verbatim}
As before, we do not introduce any primitives.
The parameter {\tt sigma} entering the initial data (\ref{eq::EMID}) 
only has to be known by the setter thorn. It is therefore given as
the named argument {\tt RealParameters} to the function 
{\tt CreateSetterThorn}.

We define two groups of variables to be set by the evaluator thorn
\begin{verbatim}
EvaluationGroups = {{"Constraints", {CEl, CB}},
                    {"endens", {rho}}};
\end{verbatim}
and correspondingly two calculations that are responsible for 
evaluating the constraints and the energy density.

\subsection{The Einstein Equations in ADM form}
\label{ADM}
As a last example we consider the Einstein equations
in ADM form, which provide a nontrivial example that 
illustrates the advantages of automatic code generation. A mere 200 lines
of Mathematica input is needed to generate a whole evolution code
including the evaluation of the constraints.
Compared to the previous example, we make use of more features of
TensorTools, for example defining a covariant derivative and
dealing with tensors with symmetries. The code generation part
illustrates the concept of primitive grid functions and the possibility of
generating two different setter thorns to compute the same quantity 
in different ways. Further, it explains how grid functions 
are computed that depend on derivatives of other grid functions. 

\subsubsection{Continuum Formulation}

We decompose the four dimensional vacuum Einstein equations 
$G_{\mu\nu} = 0$ into a set of evolution equations and constraints 
corresponding to the standard ADM formulation given by York \cite{York79}.
We will not touch upon issues of
well-posedness, as these are irrelevant for our current purpose.
Space-time is foliated by a family of spacelike hypersurfaces 
$\Sigma_{t}$, labelled by a parameter $t$. Introducing coordinates $(t, x^i)$ 
adapted to this foliation,
the four dimensional metric can be written as
\be
d s^2 = -(\alpha^2 - \beta_i \beta^i) d t^2 - 2 \beta_i dt dx^i + h_{i j} dx^i
dx^j,
\ee
where $h_{ij}$ is the induced metric on $\Sigma_t$,
$\alpha$ is the lapse function and 
$\beta^i$ the shift vector and where we set $\beta_i = h_{ij} \beta^j$.
Lapse and shift describe the components of the time flow $\partial_t$ 
orthogonal and parallel to the hypersurface $\Sigma_t$, 
$\partial_t = \alpha n + \beta^i \partial_i$. 
The rate of change of the induced metric $h_{ij}$ 
along the normal direction $n$ 
defines the extrinsic curvature $K_{ij} = \frac{1}{2}\mathcal L_{n} h_{ij}$.

Projections of the vacuum Einstein equations in the form
${}^{\scriptscriptstyle(4)}R_{\mu\nu} = 0$
into the hypersurface $\Sigma_t$, together with the definition
of the extrinsic curvature, yield evolution equations 
for $h_{ij}$ and $K_{ij}$,  
\bea\label{eq::ADMEvolEqs}
\partial_t h_{ij} & = & \mathcal L_\beta h_{ij} + 2 \, \alpha \, K_{ij} \\
\partial_t K_{ij} & = & \mathcal L_\beta K_{ij} +  D_i D_j \alpha + 
                     2 \, \alpha \, h^{lm} K_{lj} K_{mi} - \alpha \, K \, K_{ij} -
                     \alpha \, R_{i j}. \label{eq::dotKij}
\eea
Projection of $G_{\mu\nu} = 0$ in the orthogonal direction 
yields the Hamiltonian and momentum constraints,
\bea\label{eq::ADMConstrEqs}
C_H  & \equiv & R  - h^{lm} h^{pq} K_{l p} K_{m q} +  K^2   =  0 \nonumber \\
{C_M}_i & \equiv & h^{l m} D_l K_{i m} - D_i K   =  0.
\eea
$\mathcal L_{\beta}$ denotes the Lie derivative in the direction
of the shift vector $\beta^i$ and $D_i$ is the covariant
derivative compatible with the 3-metric $h_{ij}$.
The three dimensional scalar curvature $R$, Christoffel
symbols $\gamma^i{}_{jk}$ and the trace
of the extrinsic curvature $K$ have been introduced as abbreviations,
\bea\label{eq::ADMabbrev}
K & = & h^{ij} K_{ij}, \nonumber\\
\gamma^{i}{}_{jk} & = & h^{il}(\partial_k h_{lj} + 
   \partial_j h_{kl}  - \partial_l h_{jk})/2, \nonumber\\
R & = & h^{ij} R_{ij}.
\eea
Again the constraints (\ref{eq::ADMConstrEqs}) are propagated by the evolution
equations (\ref{eq::ADMEvolEqs}) and it is sufficient to
evolve freely (i.e.~without imposing the constraints at each time step) initial data
satisfying (\ref{eq::ADMEvolEqs}).

We express the Ricci tensor $R_{ij}$ in two ways, one involving
second derivatives of the metric
\be\label{eq::RiccifromMetric}   
R_{ij}  =  h^{l m} (-\partial_{i}\partial_j h_{lm} + 
         \partial_{j}\partial_m h_{l i} + \partial_i \partial_l h_{m j} - 
         \partial_{l} \partial_m h_{i j})/2 +
        h_{l m}h^{p q}
          (\gamma^l{}_{m i} \gamma^p{}_{q j} - \gamma^l{}_{ij} 
\gamma^p{}_{m q} ),
\ee
and the other is the standard representation in terms of Christoffel symbols
\be\label{eq::RicciStandard}
R_{ij} = \partial_p \gamma^p{}_{i j} - \partial_j \gamma^p{}_{p i} 
         + \gamma^p{}_{ij} \gamma^q{}_{pq} - \gamma^p{}_{q i} \gamma^q{}_{p j}.
\ee
Note that Eqs.~(\ref{eq::RiccifromMetric}) and (\ref{eq::RicciStandard})
lead to different stencil sizes when discretized, which might change the
behaviour of the numerical evolution code.

For our choice of gauge, initial data and boundary conditions,
we follow the ``gauge wave testbed'' described in \cite{mexico1}.
The spatial domain is again taken to be a 3-torus $\mathcal T$ and 
initial data are given by
\bea\label{eq::ADMGaugeWaveID}
h_{11}(0,x^i) & = & 1 - A \sin (\frac{2 \pi x}{d}), \nonumber\\
K_{11}(0,x^i) & = &  \frac{\pi A}{d}\frac{\cos(\frac{2 \pi x}{d})}
             {\sqrt{1 - A \sin(\frac{2\pi x}{d})}}
\eea
with all other components of the extrinsic curvature and all off-diagonal 3-metric components vanishing
initially. The remaining diagonal metric components are $1$ initially. 
$A$ is a constant amplitude and $d$ again denotes the range of the
coordinate $x$. The shift $\beta^i$ is set to zero and the lapse $\alpha$ 
is given by the square root of the determinant of the 3-metric 
$\alpha = \sqrt{\textrm{det} h}$.

The analytical solution to this initial value problem is a pure gauge wave
travelling in the $x$ direction at the speed of light, $h_{11}(t,x^i) =
h_{11}(x-t), \, K_{11}(t,x^i) = K_{11}(x-t)$, where all other evolution
variables maintain their initial values.

\subsubsection{Specifications for Kranc}

The conversion of tensorial equations to equations for individual
components using the package TensorTools clearly is more involved
than in the previous examples. We start by defining the tensors appearing
in Eqs.(\ref{eq::ADMEvolEqs}) -- (\ref{eq::ADMabbrev})  
\begin{verbatim}
Map[DefineTensor,{h, hInv, K, beta, Ricci, gamma, CM}]
\end{verbatim}
The inverse metric $h^{ij}$ has been renamed to {\tt hInv} as it is going to 
be a shorthand computed from the metric $h_{ij}$. {\tt Ricci} and {\tt gamma}
denote the Ricci tensor and the Christoffel symbols respectively, and {\tt beta}
is the shift.

Note that since the tensors $h_{ij}$ and $K_{ij}$ are symmetric, we
only evolve a subset of linearly independent components. To this effect, 
before applying {\tt MakeExplicit} to expressions involving symmetric
tensors, the symmetries are registered. Note that, in conformance
with the conventions of MathTensor,  in our example covariant and
contravariant indices are treated differently with respect to the 
ordering of coordinate indices. The ``{\tt 12}'' component
of a covariant symmetric 2-tensor automatically is converted to the 
``{\tt 21}'' component, whereas for contravariant symmetric tensors the
index ordering is the opposite. The corresponding commands for symmetric
(co/contravariant) tensors are {\tt AssertSymmetricDecreasing} and 
{\tt AssertSymmetricIncreasing},
\begin{verbatim}
AssertSymmetricIncreasing[hInv[ua,ub]];
Map[AssertSymmetricDecreasing, {h[la,lb], K[la,lb], Ricci[la,lb]}];
AssertSymmetricDecreasing[gamma[ua, lb, lc], lb, lc];
\end{verbatim}
The symmetry handling in TensorTools is rudimentary, and currently
only supports symmetries for pairs of indices in any particular
tensor.  When applying {\tt MakeExplicit} to one of the above tensors all
components are kept thus resulting in a list with duplicate entries.
These have to be removed by a user defined function which is called {\tt
MakeExplicitSet} here.

To complete the description of the geometric structure for TensorTools
the covariant derivative together with the corresponding
Christoffel symbol is defined,
\begin{verbatim}
DefineConnection[CD, gamma];
\end{verbatim}

Computing the determinant of the spatial metric $h$ by the Mathematica component expression
{\tt hDet = Det[MatrixOfComponents[h[la,lb]]]}
the inverse metric is given by a series of shorthands
\begin{verbatim}
deth        -> hDet,
invdeth     -> 1 / deth,
hInv[ua,ub] -> invdeth hDet MatrixInverse[h[ua,ub]]
\end{verbatim}
Here, {\tt deth} is computed as a shorthand at run time to avoid it
being calculated for each component of {\tt hInv[ua,ub]}.  {\tt hDet
MatrixInverse[h[ua,ub]]} is the component inverse of {\tt h} with the
common determinant factor cancelled.

The quantities appearing in Eqs.
(\ref{eq::ADMEvolEqs}) and (\ref{eq::ADMConstrEqs}) split
in the following way into evolution variables, primitives and shorthands:
Evolution variables are the 3-metric and the extrinsic curvature
\begin{verbatim}
ADMvars = {{"h", {h11, h21, h31, h22, h32, h33}}, {"K",{K11, K21, K31,K22, K32, K33}}}};
\end{verbatim}
All other quantities appearing in (\ref{eq::ADMEvolEqs}) and
(\ref{eq::ADMConstrEqs}) have to be either primitive grid functions, 
which are set in a setter thorn, or shorthands defined inside the loop in which they
are used. The decision about which quantities are primitives is based on the
following considerations. Quantities that appear in differentiated form in
the evolution equations, e.g.~lapse and shift, have to be grid functions. 
Quantities which can be computed in various different ways, 
e.g.~the Ricci tensor,
can be set in different setter thorns which then can be 
compiled into the same executable, 
where one of them is ``activated'' at run time. Quantities that are needed
for both the evolution equations and the constraints, e.g.~the Christoffel
symbols, can be set in a setter thorn in order to reduce computational costs.
Furthermore all quantities whose output is required, e.g.~the trace of the
extrinsic curvature as specified in the tests in \cite{mexico1}, 
have to be grid functions.  All of these quantities should be declared as primitives. 
The complete list of primitives for our example therefore is
\begin{verbatim}
ADMprimitives = {{"alpha",{alpha}}, {"beta",{beta1, ...}}, {"Ricci",{Ricci11, ...}},
                 {"gamma",{gamma111, ...}}, {"trK",{trK}}} 
\end{verbatim}
The scalar curvature is computed as a shorthand from the Ricci tensor. 

Eqs.~(\ref{eq::ADMEvolEqs}) -- (\ref{eq::RicciStandard}) in MathTensor
syntax are read from the file {\tt ADMEqs.m} and expanded into components 
with {\tt MakeExplicit}.
In order to increase performance
once the functionality of TensorTools is not needed anymore we
switch off the special treatment of products by setting
\begin{verbatim}
SetEnhancedTimes[False]
\end{verbatim}
As a last preparational step we define the components of the inverse metric
and the lapse in a {\tt CollectList} which is going to be used to 
simplify the equations at the level of components, essentially
by applying the Mathematica function 
{\tt Collect[rhs, CollectList, Simplify]} to the right hand sides
of assignments.

The generation of the base thorn, the MoL thorn, the various setter thorns
and the evaluator thorn run along the same lines as in the previous
examples with the following additional features.  All primitive grid
functions have to be set at every time step, so the argument {\tt SetTime}
for all setter thorns therefore has to be {\tt "initial\_and\_poststep"}.

The Ricci tensor in the form Eq.~(\ref{eq::RicciStandard}) requires taking 
derivatives of the Christoffel symbols, which therefore have to be known 
on the whole grid before the Ricci tensor is computed.
This is done in the code by using two loops, where the first loop
is used to compute the Christoffel symbols. At the end of the loop
the grid functions that store the Christoffel symbols 
are synchronized such that the ghost zones are filled in with data from the opposite 
side of the grid.
The Ricci tensor can then be computed in the second loop.
In order to generate two separate loops,
the calculation passed as an argument to the function CreateSetterThorn
has to specify the equations in two separate (sub)lists, i.e.
\begin{verbatim}
Equations -> {Join[InvMetricEqs, ChristoffelEqs], RicciStandardEqs};
\end{verbatim}
 
Another major difference to the previous examples is the way the initial
data is set. We use the existing thorn {\tt Exact} in the
arrangement {\tt AEIThorns} 
to compute the metric and
extrinsic curvature as given in Eq.~(\ref{eq::ADMGaugeWaveID}). The only thing left to do
is to convert the initial data from the variables used by the Exact thorn, 
{\tt \{gxx, ..., kxx,....\}}, defined by the {\tt CactusEinstein} arrangement, to our evolution variables and
correct for different sign conventions in the definition of the extrinsic
curvature, i.e.~{\tt h11 == gxx, ..., K11 == - kxx, ...}. This is done in the
Translator thorn, which implements functions to translate either way:
from the 
{\tt CactusEinstein} variables to the evolution variables at the 
{\tt POSTINITIAL} Cactus time bin, and from the evolution variables to the {\tt CactusEinstein}
variables after every time step. 
The latter translation is 
performed to allow use of existing analysis
tools,  e.g.~in {\tt CactusEinstein}.
The translator functions are defined by the Kranc
calculations {\tt TranslatorInCalculation} and
{\tt TranslatorOutCalculation} respectively. 

The thorn list is generated as in the other examples. This time, however,
it has to be supplemented by the thorns {\tt CactusEinstein/ADMCoupling,
AEIThorns/Exact, CactusEinstein/CoordGauge} and 
{\tt CactusEinstein/StaticConformal} which are needed by the Exact thorn to set
up the initial data Eqs.~(\ref{eq::ADMGaugeWaveID}). 

All the thorns contained in the thorn list -- including both 
Setter thorns for the Ricci tensor -- are compiled into one executable.
The parameter file specifies the way the Ricci tensor is computed 
by activating the relevant thorn. The specification of the
parameters determining the initial data can be found in the documentation
of the thorn {\tt AEIThorns/Exact}.

\subsubsection{Performance}
\label{sec:Performance}

We include some timing results for simulations
to demonstrate that our computer-generated code can
achieve a performance that is comparable to carefully optimized handwritten 
code.
We do not attempt a systematic study of performance issues, and restrict
ourselves to a single example based on the ``gauge wave'' data as
defined in Eq.~(\ref{eq::ADMGaugeWaveID}).
Our test run is a 3D version of the gauge wave spacetime defined
with a grid size of $50^3$ (plus one ghost zone); we evolve for 100 time steps.

We compare the performance of the ADM code presented above
to a code that has been developed as a production code by the AEI
astrophysical numerical relativity group. This code is publicly available
from the Cactus CVS server as the CactusEinstein arrangement 
\cite{CactusEinstein}.
The Kranc and CactusEinstein ADM codes use the same numerical methods 
(3 step Iterative Crank Nicolson and second order centred spatial finite 
differencing).
As a physics production code, CactusEinstein ADM is significantly
more complex, but the evolution algorithm is essentially
equivalent. The numbers given below should only be taken as a rough guide. 
In contrast to Kranc ADM, which uses computer generated code and a generic 
infrastructure for spatial differencing and time evolution,
CactusEinstein ADM is hand coded and does not support general numerical schemes
(although different time update schemes have been implemented as well).

In the tables \ref{laptopResults} and \ref{workstationResults},
 {\em compile time} and {\em user runtime} have been
timed with Linux's time utility and show the total number of CPU-seconds that
the process spent in user mode.  The results of the {\em cactus timing} column
have been obtained with the internal timing infrastructure of Cactus.

Tests have been performed on two machines, a Dell D600 Laptop running Redhat
Linux 9.0, with a 1.6 GHz Intel Pentium M CPU and 1 GByte of RAM;
and a Dell precision 450 workstation running Redhat Linux 7.2, with a dual 
Intel Pentium Xeon CPU running at 1.7 GHz and again 1 GByte of RAM.
All runs have been performed as single processor runs.

Note that with the exception of the result for CactusEinstein ADM compiled
with Intel 7.1, the Kranc generated code is around 6 -- 10 \% slower.
Possible explanations are that the CactusEinstein ADM code partially
uses Fortran 90 array language which allows further optimizations
(e.g.~the vectorization feature of the Intel compiler), and
the scheme for synchronizing grid functions within Kranc is not yet optimal 
(synchronizing is not necessary if no derivatives appear in assignments, as the
ghost zones can be updated locally).
We speculate that the CactusEinstein result for the Intel 7.1 compiler might be
caused by the compiler reaching internal limits and giving up on certain kinds
of optimization.
\begin{center}
\begin{table}
\label{laptopResults}
\begin{tabular}{|l|r|r|r|}\hline\hline
  code version                        & compile time & user runtime & cactus timing \\\hline
  CactusEinstein ADM, Intel 8         & 184          & 145.2 & 145.6  \\
  CactusEinstein ADM, VAST f90 \& gcc & 154          & 155.0 & 155.2  \\
  Kranc ADM, C,   Intel 8             & 119          & 154.9 & 155.4  \\
  Kranc ADM, C,   VAST f90 \& gcc     &  89          & 166.8 & 167.4  \\
  Kranc ADM, F90, Intel 8             & 129          & 159.7 & 159.5  \\
  Kranc ADM, F90, VAST f90 \& gcc     &  92          & 168.3 & 168.8  \\
  \hline\hline
\end{tabular}
\caption{Timing results from a Dell D600 Laptop running Redhat Linux 9.0, with a 1.6 GHz Intel Pentium M CPU and 1GByte of RAM, comparing results obtained with the Intel 8 compilers vs. a combination of Pacific Sierra VAST F90 and gcc 3.2.2 compilers.  Optimization options for the Intel compilers are 
{\tt -O3 -xN -ip}, and {\tt -O3} for VAST and gcc.}
\end{table}
\end{center}
\begin{center}
\begin{table}
\label{workstationResults}
\begin{tabular}{|l|r|r|r|}\hline\hline
  code version                    & compile time & user runtime & cactus timing \\\hline
  CactusEinstein ADM, Intel 7.1   & 303          & 188.1 & 189.5  \\
  Kranc ADM, C, Intel 7.1         & 186          & 131.3 & 131.1  \\
  Kranc ADM, F90, Intel 7.1       & 194          & 115.3 & 115.7  \\
  \hline\hline
\end{tabular}
\caption{Timing results from a Dell Precision 450 workstation running Redhat Linux 7.2, with a 1.7 GHz Intel Pentium Xeon CPU and 1GByte of RAM, binaries have been built with the Intel 7.1 compilers with optimization options 
{\tt-O3 -xW -ip}.}
\end{table}
\end{center}

\begin{acknowledgments}
The authors thank the members of the AEI numerical relativity group
and the Cactus developers for helpful discussions, 
in particular D. Pollney and I. Hawke for many stimulating discussions, 
I. Hawke for helping with MoL,
E. Schnetter for resolving subtleties with preprocessing and compiling
Kranc-generated code, J. Thornburg for careful reading of an earlier version
of the manuscript, and K. Roszkowski and N. Dorband for feedback about using
Kranc. I. Hinder thanks AEI for hospitality.
\end{acknowledgments}

\appendix
\section{Data structure specifications}
\label{app:data_structures}

Here we describe in detail the data structures which are used when
calling the KrancThorns functions.

\subsection{Calculation}
\label{app:Calculation}

\begin{center}
\begin{tabular}{|l|l|p{3in}|}
  \hline
  \bf Key & \bf Type & \bf Description \\
  \hline
  Equations       & list of lists    & {\tt \{loop1, loop2\}} -- Each loop is a list
                                       of rules of the form {\tt {\it variable} -> 
                                       {\it expression}} where
                              {\tt \it variable} is to be set from {\tt \it expression} \\
  Shorthands (optional)  & list of symbols    & Variables which are to be considered
                              as `shorthands' for the purposes of this calculation \\
  Name (optional) & string  & A name for the calculation \\
  Before (optional) & list of strings  & Function names before which the
                              calculation should be scheduled. \\
  After (optional)  & list of strings &  Function names after which the
                              calculation should be scheduled. \\
  \hline
\end{tabular}
\end{center}

\subsection{GroupCalculation}
\label{app:GroupCalculation}

A GroupCalculation structure is a list of two elements; the first is
the name (a string) of a Cactus group and the second is the
Calculation to update the variables in that group.

\subsection{GroupDefinition}
\label{GroupDefinition}

A GroupDefinition structure is a list of two elements.  The first is
the name (string) of a Cactus group and the second is the list of
variables (symbols) belonging to that group. The group name can be
prefixed with the name of the Cactus implementation that provides the
group followed by two colons (e.g.~``ADMBase::metric'').  If this is
not done, then the KrancThorns functions will attempt to guess the
implementation name, usually using the name of the thorn being
created.

\section{KrancThorns function reference}
\label{app:function_reference}

Here we document the arguments which can be specified for the
functions CreateBaseThorn, CreateMoLThorn, CreateSetterThorn,
CreateTranslatorThorn and CreateEvaluatorThorn.

Note that we use Mathematica syntax for function-specific section headers.
Underscores denote function arguments, and OptArguments stands for optional
arguments, also referred to as named arguments below.  These
are given in the form {\tt myFunction[..., argumentName -> argumentValue]}.


\subsection{Common Named Arguments}
\label{app:common_arguments}

The following named arguments can be used in any of the Create*Thorn
functions:
\begin{center}
\begin{tabular}{|l|l|p{3in}|l|}
  \hline
  \bf Argument & \bf Type & \bf Description & \bf Default\\
  \hline
  SystemName & string & A name for the evolution system implemented by this arrangement.
                        This will be used for the name of the arrangement directory
                          & ``MyPDESystem''\\
  SystemParentDirectory & string & The directory in which to create the arrangement directory & ``.''\\
  ThornName & string & The name to give this thorn & SystemName + thorn type\\
  Implementation & string & The name of the Cactus implementation that this thorn defines & ThornName\\
  SystemDescription & string & A short description of the system implemented by this arrangement & SystemName \\
  DeBug & Boolean & Whether or not to print debugging information during thorn generation& False\\
  \hline
\end{tabular}
\end{center}

\subsection{Arguments relating to parameters}
\label{app:parameter_arguments}

The following table describes named arguments that can be specified
for any of the thorns except CreateBaseThorn.  CreateBaseThorn is
special because it can be used to define parameters which are
inherited by each thorn in the arrangement, so the arguments it can be
given are slightly different.

\begin{center}
\begin{tabular}{|l|l|p{3in}|l|}
  \hline
  \bf Argument & \bf Type & \bf Description & \bf Default\\
  \hline
  RealBaseParameters & list of strings & Real parameters used in this thorn but 
                                         defined in the base thorn & \{\}\\
  IntBaseParameters & list of strings & Integer parameters used in this thorn 
                                        but defined in the base thorn & \{\}\\
  RealParameters & list of strings & Real parameters defined in this thorn & \{\}\\
  IntParameters & list of strings & Integer parameters defined in this thorn & \{\}\\
  \hline
\end{tabular}
\end{center}


\subsection{CreateBaseThorn[groups\_, evolvedGroupNames\_, primitiveGroupNames\_, 
OptArguments\_\_\_]}
\label{app:CreateBaseThorn}

\subsubsection{Positional arguments}

\begin{center}
\begin{tabular}{|l|l|p{3in}|}
  \hline
  \bf Argument & \bf Type & \bf Description \\
  \hline
  groups              & list of GroupDefinition structures &

  Definitions of any groups referred to in the other arguments.  Can
  supply extra definitions for other groups which will be safely
  ignored.  \\

  evolvedGroupNames & list of strings &
  
  Names of groups containing grid functions which will be evolved by
  MoL in any of the thorns in the arrangement. \\

  primitiveGroupNames & list of strings &

  Names of groups containing grid functions which will be referred to
  during calculation of the MoL right hand sides in any of the thorns
  in the arrangement. \\

  \hline
\end{tabular}
\end{center}

\subsubsection{Named arguments}

\begin{center}
\begin{tabular}{|l|l|p{3in}|l|}
  \hline
  \bf Argument & \bf Type & \bf Description & \bf Default\\
  \hline
  RealBaseParameters & list of strings & Real parameters defined in this thorn 
and inherited by all the thorns in the arrangement & \{\}\\
  IntBaseParameters & list of strings & Integer parameters defined in this thorn 
and inherited by all the thorns in the arrangement& \{\}\\
  \hline
\end{tabular}
\end{center}


\subsection{CreateEvaluatorThorn[groupCalculations\_, groups\_, OptArguments\_\_\_]}
\label{app:CreateEvaluatorThorn}

\subsubsection{Positional arguments}

\begin{center}
\begin{tabular}{|l|l|p{3in}|}
  \hline
  \bf Argument & \bf Type & \bf Description \\
  \hline
  groupCalculations
  & list of GroupCalculation structures 
  & The GroupCalculations to evaluate in order to
    set the variables in each group
  \\
  groups 
  & list of GroupDefinition structures 
  & Definitions for each of the groups referred to in this thorn. Can
  supply extra definitions for other groups which will be safely
  ignored. \\

  \hline
\end{tabular}
\end{center}


\subsection{CreateMoLThorn[calculation\_, groups\_, OptArguments\_\_\_]}
\label{app:CreateMoLThorn}

\subsubsection{Positional Arguments}

\begin{center}
\begin{tabular}{|l|l|p{3in}|}
  \hline
  \bf Argument & \bf Type & \bf Description \\
  \hline
  calculation & Calculation & The calculation for setting the right hand side variables
                              for MoL.  The equations should be of the form
                              {\tt dot[{\it gf}] -> {\it expression}} for evolution equations, and
                              {\tt {\it shorthand} -> {\it expression}} for shorthand definitions, which can
                              be freely mixed in to the list.\\
  groups & list of GroupDefinition structures 

& Definitions for each of the groups referred to in this thorn. Can
  supply extra definitions for other groups which will be safely
  ignored. \\
  \hline
\end{tabular}
\end{center}

\subsubsection{Named Arguments}

\begin{center}
\begin{tabular}{|l|l|p{3in}|l|}
  \hline
  \bf Argument & \bf Type & \bf Description & \bf Default\\
  \hline
  PrimitiveGroups & list of strings & These are the groups containing the grid functions which are
                                      referred to but not evolved by this evolution thorn & \{\}  \\
  \hline
\end{tabular}
\end{center}


\subsection{CreateSetterThorn[calculation\_, OptArguments\_\_\_]}
\label{app:CreateSetterThorn}

\subsubsection{Positional Arguments}

\begin{center}
\begin{tabular}{|l|l|p{3in}|}
  \hline
  \bf Argument & \bf Type & \bf Description \\
  \hline
  calculation & Calculation & The calculation to be performed \\

  groups & list of GroupDefinition structures 

& Definitions for each of the groups referred to in this thorn. Can
  supply extra definitions for other groups which will be safely
  ignored. \\

  \hline
\end{tabular}
\end{center}

\subsubsection{Named Arguments}

\begin{center}
\begin{tabular}{|l|l|p{3in}|l|}
  \hline
  \bf Argument & \bf Type & \bf Description & \bf Default\\
  \hline
  SetTime (optional) & string & ``initial\_and\_poststep'', 
                     ``initial\_only''
                     or ``poststep\_only'' & ``initial\_and\_poststep'' \\
  \hline
\end{tabular}
\end{center}


\subsection{CreateTranslatorThorn[groups\_, OptArguments\_\_\_]}
\label{app:CreateTranslatorThorn}

\subsubsection{Positional Arguments}

\begin{center}
\begin{tabular}{|l|l|p{3in}|}
  \hline
  \bf Argument & \bf Type & \bf Description \\
  \hline

  groups & list of GroupDefinition structures 

& Definitions for each of the groups referred to in this thorn. Can
  supply extra definitions for other groups which will be safely
  ignored. \\
  \hline
\end{tabular}
\end{center}

\subsubsection{Named Arguments}

\begin{center}
\begin{tabular}{|l|l|p{3in}|}
  \hline
  \bf Argument & \bf Type & \bf Description\\
  \hline
  TranslatorInCalculation & Calculation & The calculation to set the evolved 
                                          variables from some other source  \\
  TranslatorOutCalculation & Calculation & The calculation to convert the evolved
                                           variables back into some other set of 
                                           variables  \\
  \hline
\end{tabular}
\end{center}



\begin{thebibliography}{99}

\bibitem{Luis}
For an overview of current issues and PDE problems arising
in numerical relativity, and for further references, see, e.g,~L.\ Lehner,
{\it Class.\ Quant.\ Grav.\/} {\bf 18}, R25 (2001).


\bibitem{EriksCode} Erik Schnetter, Gauge fixing for the simulation of black hole spacetimes, 
PhD thesis, Universit\"at T\"ubingen, 2003 
({\tt http://w210.ub.uni-tuebingen.de/dbt/volltexte/2003/819/}).

\bibitem{grtensor}
 P. Musgrave, D. Pollney and K. Lake,
 {\it Fields Institute Comm.} {\bf 15} 313 (1996) ({\tt http://grtensor.phy.queensu.ca}).

\bibitem{Wald}
 R. Wald, {\sl General Relativity}, University of Chicago Press 1984.

\bibitem{PenroseRindler1984} 
 R. Penrose and W. Rindler,
 {\sl Spinors \& Space-time: Two-spinor Calculus \& Relativistic Fields},
 Cambridge University Press 1984.

\bibitem{Ricci}
 J. Lee,
 {\sl Ricci, a Mathematica package for doing tensor calculations
in differential geometry},
{\tt http://www.math.washington.edu/$\sim$lee}.

\bibitem{xTensor}
 J. M. Mart\'{\i}n-Garc\'{\i}a,  
 {\sl xTensor}, {\tt http://metric.imaff.csic.es/Martin-Garcia/xAct/}.

\bibitem{MathTensor}
 L. Parker and S. Christensen,
 {\sl MathTensor:  A System for Doing Tensor Analysis by Computer},
 Addison-Wesley 1994.

\bibitem{MPI}
 The MPI Forum,
 {\tt http://www-unix.mcs.anl.gov/mpi}.


\bibitem{Cactus}
  T. Goodale, G. Allen, G. Lanfermann, J. Mass\'o, T. Radke, E. Seidel and J. Shalf,
        "The {C}actus Framework and Toolkit: Design and Applications", in
    {\em Vector and Parallel Processing - VECPAR'2002, 5th International Conference,
     Lecture Notes in Computer Science}, Springer, Berlin, 2003;
 G. Allen et. al., {\it Cluster Computing}, {\bf 4} 179 (2001);
B. Talbot, S. Zhou, G. Higgins, {\it NASA report},
{\tt http://ct.gsfc.nasa.gov/esmf\_tasc/Files/Cactus\_b.html};
Cactus development team,  http://www.cactuscode.org.


\bibitem{PETSc} 
{\sl Portable, Extensible Toolkit for Scientific Computation},
{\tt http://www-unix.mcs.anl.gov/petsc/petsc-2}.

\bibitem{JonathanAH}
 J. Thornburg,
 {\it Class.\ Quant.\ Grav.\/} {\bf 21}(2), 743-766 (2004).

\bibitem{PeterEH}
 P. Diener,
 {\it Class.\ Quant.\ Grav.\/} {\bf 20}, 4901-4918 (2003).

\bibitem{mol}
 B. Gustafsson, H. Kreiss and J. Oliger,
 {\sl Time Dependent Problems and Difference Methods},
 John Wiley and Sons 1995.

\bibitem{CactusMoL} I. Hawke, for a short description and download instructions
see {\tt http://www.aei.mpg.de/$\sim$hawke/MoL/MoL.html}.


\bibitem{cartoon_paper}  M. Alcubierre, S. Brandt, B. Bruegmann, D. Holz, 
E. Seidel, R. Takahashi, J. Thornburg,  Int. J. Mod. Phys. D 10 (2001)
273--290.

\bibitem{ERE}
S. Husa and C. Lechner, 
in Proceedings of the Spanish Relativity Meeting (ERE 2002),
Ma\'o, Menorca, Spain, 22--24 Sept 2002, ed. by A. Lobo et al., Barcelona
University Press 2003, e-Print archive gr-qc/0301076.

\bibitem{York79}
J.W. York, Jr., `Kinematics and Dynamics of General Relativity", 
in {\em Sources of Gravitational Radiation}, ed. by L.Smarr, 
(Cambridge, 1979).

\bibitem{mexico1}
M.~Alcubierre et al, {\it Class.\ Quant.\ Grav.\/} {\bf 21}, 
589-613 (2004).

\bibitem{CactusEinstein}
See
{\tt http://www.cactuscode.org/Community/Relativity.html}.
\end{thebibliography}
\end{document}